\documentstyle[11pt,paspconf]{article}
\input{epsf}
\def\edcomment#1{\iffalse\marginpar{\raggedright\sl#1\/}\else\relax\fi}
\reversemarginpar
\begin{document}
\title{The Cluster Temperature Function at High Redshift}
\author{Paolo Tozzi}
\affil{II Univ. di Roma, Tor Vergata -- tozzi@axtov1.roma2.infn.it}
\author{Fabio Governato}
\affil{Durham University -- Fabio.Governato@durham.ac.uk}

\begin{abstract}

We take advantage of the biggest cosmological simulation to date for a
critical CDM universe in order to test robustness of the cluster mass
function on a range of masses much wider than tested before.  On the high
mass end our results show an excess of hot clusters and a milder evolution
compared to the analytical predictions based on the Press \& Schechter
formula.  These features must be properly taken into account in deriving
the cluster X-ray temperature function at moderate and high redshifts.  On
a general basis, the reduced negative evolution in the number of hot
clusters could alleviate the discrepancies between the predictions for
critical universes and incoming data, which instead seem to favour low
$\Omega_0$.
\end{abstract}

\section{Introduction}

The crucial link between the initial density perturbations in the linear stage 
and the present day high contrast objects is made by the non linear 
gravitational dynamics.  This can be followed in full 
detail only by very time consuming N--body simulations.  To achieve this, 
several parallel N--body algorithms have been developed recently.  
% A large consensus has been  reached 
% on the results obtained from different N--body techniques (Weinberg....1997).  
Meanwhile, 
analytical tools has been checked on these numerical results, leading to 
the wide popularity of the Press \& Schechter (1974, PS) formula as a fair 
description of both the shape and evolution of the cluster mass function
$N(M)$.  A detailed description of the PS statistics with an
understanding of its dynamical basis can be found in Bond et al. (1991).
Its main advantage
compared to other analytical methods lies in its simplicity and 
applicability to many different cosmologies.  

The agreement between PS and the N--body is, at some level, very 
surprising because the PS formula is based
entirely on the linear theory, neglecting all the non linear 
effects which are expected to be important.  
Here we just recall its essential ingredients: the dispersion $\sigma(M)$ of
the linear perturbation field; the linear growth factor $D(z)$; 
the linear collapse
threshold $\delta_{c}$.  The latter is generally set to the top--hat value 
$\delta_c=1.686$.  
% (see, e.g., Peebles 1993).  

Thus all the complex dynamical processes of formation and evolution of
virialized objects are condensed into two numbers: $\delta_c$ and $D(z)$.
In a critical universe the strong dependence $D(z)\propto (1+z)^{-1}$
reflects in a strong negative evolution at higher $z$ for $N(M)$.

% The full expression for the PS formula is: 
% $$
% N(M,z)=\sqrt{{2\over \pi}}\rho_b {{\delta_{co}}\over {D(z)\sigma (M)}}
% {{d\ln \sigma}\over {d\ln M}}{1\over {M^2}}\exp ^{{1\over 2}{{\delta^2_{co}
% \over 
% {D^2(z)\sigma^2(M)}}}}
% $$

% However, some discrepancies was enlightened in
% the past (Bertschinger \& Jain 1993), pointing at an excess in the
% high mass end of the mass distribution seen in numerical simulations
% respect to the PS predictions.  Such discrepancies are extremely relevant 
% for predictions at moderat and high redshifts.  
% Then it is extremely important to assess the amount of such discrepancies.  

To assess the reliability of the PS predictions of cluster temperature
distribution at high $z$,  we first test both shape and evolution of the mass
function in the case of a critical CDM universe.  

\section {Results for a Critical CDM}

The simulations were run using a by a parallel
treecode which allows periodic boundary conditions and individual time
steps (Quinn et al. 1997).  The volume simulated was 500$h^{-1}$Mpc
with $H_0= 50$ km/sec/Mpc, $\Omega_0=1$ and $\sigma_8=0.7$.  
We used 360$^3$ particles,  almost 47 million.  
We adopt the CDM power spectrum  given 
in Bardeen et al. (1986).

\begin{figure}
\plottwo{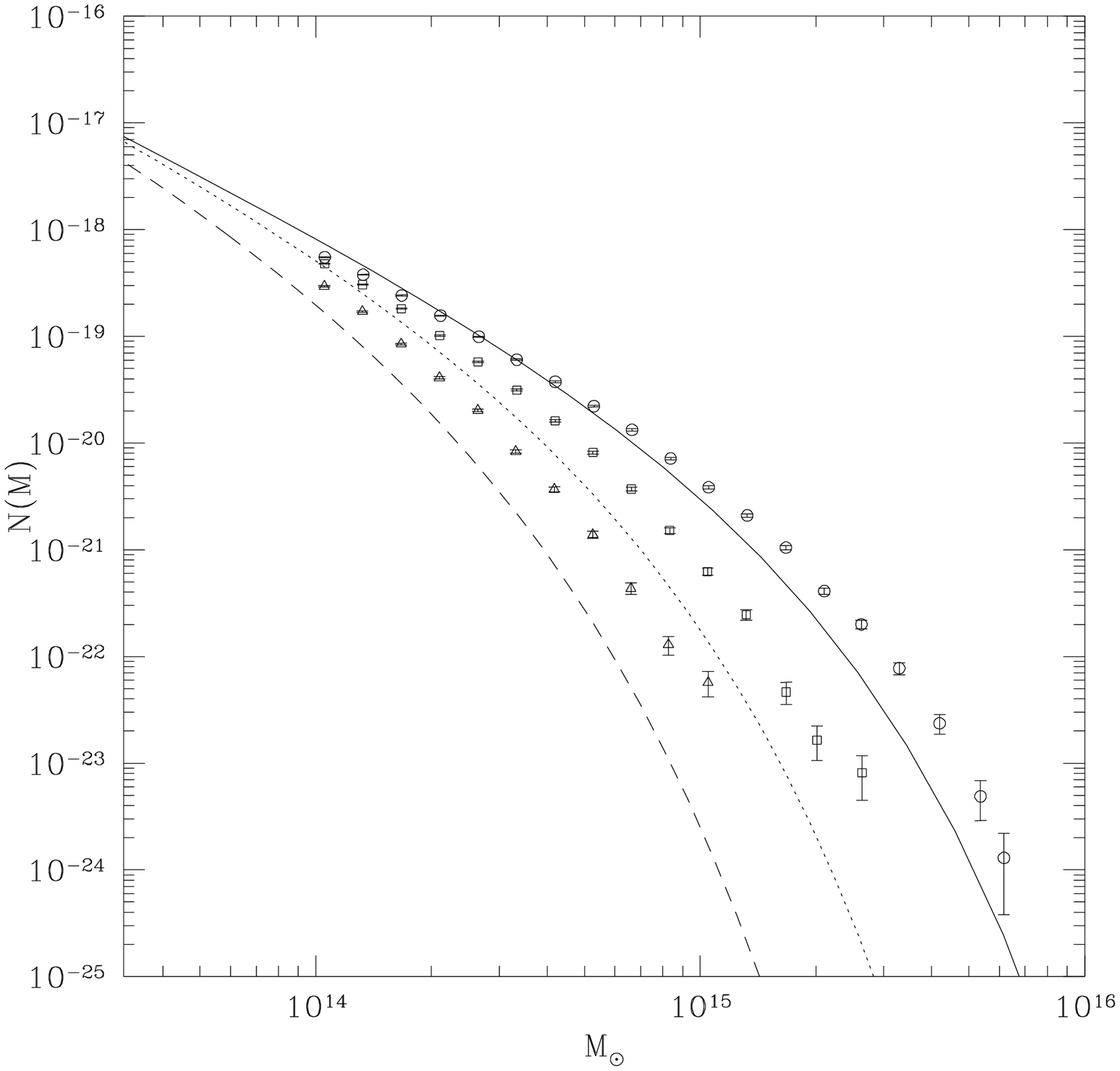}{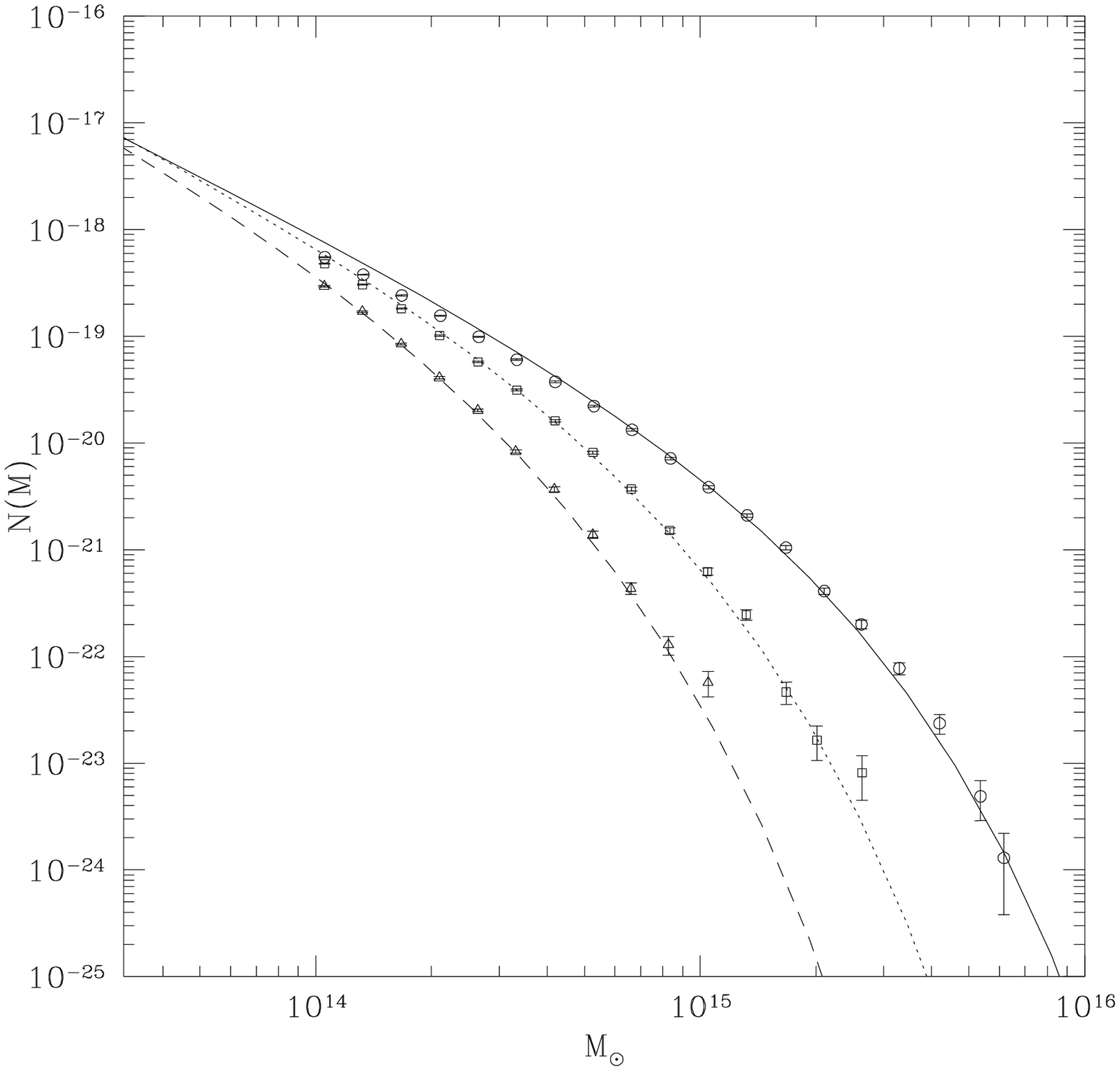}
\caption{On panel a) the N-body mass function  (points) is compared 
with the canonical PS prediction with $\delta_c=1.686$ (lines) at different 
redshifts: 0 (circles, continuous line), 0.5 (squares, dotted line), 1
(triangles, dashed line).  On panel b) the N-body mass function with 
the PS prediction after
changing $\delta_c$ to the best fit value $\delta_{eff}$ at each output 
(N(M) is in $M^{-1}_\odot$ Mpc$^{-3}$).}
\label{fig1}
\end{figure}

\begin{figure}
\plotone{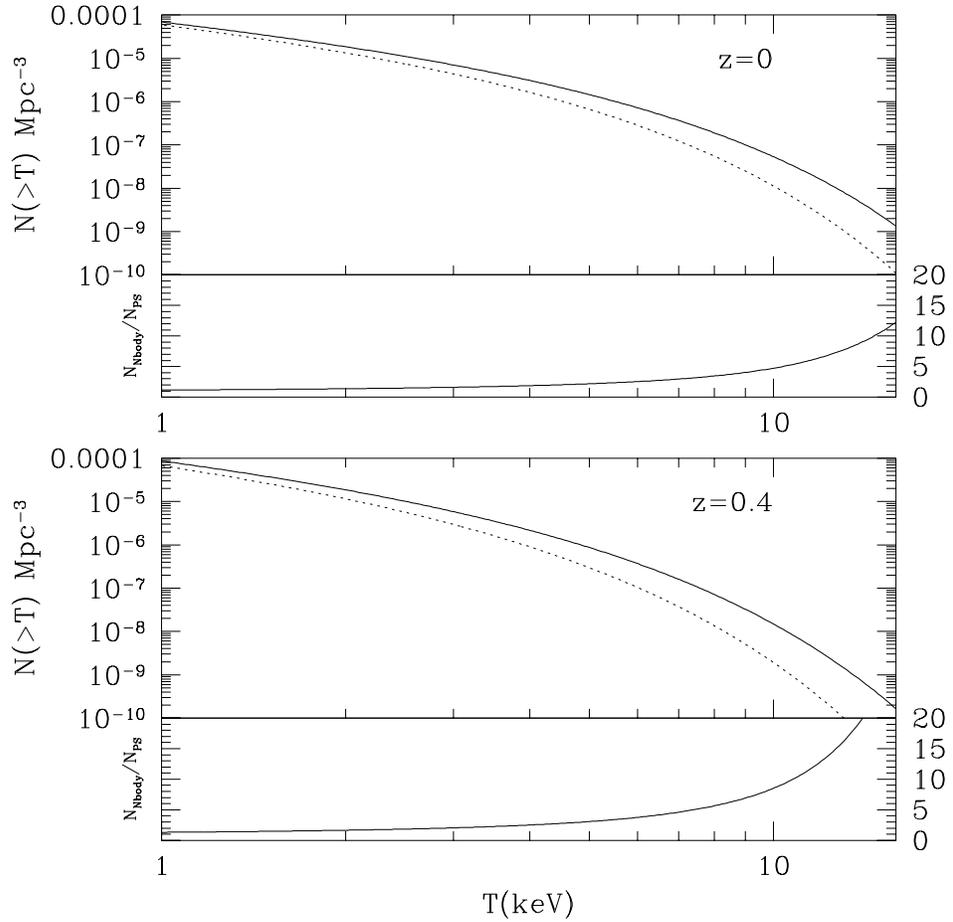}
\caption{The cumulative temperature function $N(>T)$ in a critical 
CDM universe 
at $z=0$ and $z=0.4$ (continuous lines), compared with the canonical 
PS predictions (dotted lines).  In the lower boxes, the ratio of the
N--body cluster density to PS prediction is shown.  Discrepancies become
important at $z\geq 0.4$ and $kT \geq 7$ keV.}
\label{nt}
\end{figure}

In figure \ref{fig1} a) we show the resulting mass distribution at
 three output 
redshifts $0$, $0.5$ and $1$.   The rapid evolution is 
already evident at moderate redshifts.  Clearly the local mass
function is underestimated for $M> 10^{15} M_\odot$ with respect
to the canonical ($\delta_c=1.686$) PS prediction.  The 
discrepancies grow at higher redshift, and at $z=1$ the number density
is underestimated by a factor $\geq 3$ for $M > 3 \, 10^{14}M_\odot$.  

We perform a $\chi^2$ test on the high mass end to evaluate the best fit
parameter $\delta_{eff}(z)$ at each output in redshift.  
We show that $\delta_{eff}$ is clearly always smaller than the 
top--hat value $1.686$, and it is significantly lower for higher redshift, 
though the decrement is only $6\%$ at $z\sim 1$.  
This dependence can be approximated
by a simple power law: $\delta_{eff}(z)=1.48(1+z)^{-0.06}$.  
The N-body mass function, plotted against the PS mass 
function with the best fit value $\delta_{eff}$ instead of $\delta_c$ 
at each $z$, is shown in figure \ref{fig1} b).

% \begin{figure}
% \plotone{dc.ps}
% \caption{The points shows $\delta_{eff}$ at various 
% outpts, with error bars at $2-\sigma$ confidence level.  The continuous 
% line is the power law interpolation $\delta_{eff}=1.48(1+z)^{0.06}$.  
% The dashed line is the canonical value $\delta_c=1.686$.  }
% \label{dc}
% \end{figure}

We carefully match our data with previous works, which often claimed a very
good agreement with the PS predictions (see, for a recent work, Eke et
al. 1996).  Such previous simulations explored a smaller mass range than
the one available to us, with poorer statistics and  larger
uncertainties.  On the same range and with the same statistics, we would
claim good agreement with both the analytical PS prediction and previous
numerical works.

\section{The Cluster temperature distribution} 

Under the assumption of dynamical equilibrium, the temperature 
of the intra--cluster plasma is directly linked to the depth of 
the potential wells.  The relation $T(M)$ can be easily recovered in the case of
a simple spherical collapse model (see, e.g., Eke et al. 1996).  
% for an isothermal plasma:
% $$
% k T = 5.5 M_{15}^{2/3}(1+z)\big({{\Omega_o \Delta_c}\over {\Omega(z) 178}}
% \big)
% $$
% where $\Delta_c$ is the average density contrast at virialization respect 
% to the critical background density 
This relation has been successfully tested against N--body 
hydrodynamical simulations (Navarro et al. 1995).  

The evolution of the  temperature distribution is very sensitive
to the adopted model of structures formation, and its evolution at moderate 
redshift is considered a crucial test for CDM models (for recent works in this 
field see Mathiesen \& Evrard 1997, Kitayama \& Suto 1997). 

In figure \ref{nt} the local cumulative function $N(>T)$ is plotted for the
canonical PS (dotted line) and numerical prediction (continuous line).
While discrepancies are less than a factor $3$ for $kT \leq 8$ keV, they
become significant at $z=0.4$, where the N-body simulation predicts
significant excess above $kT \simeq 7$ keV.  This not only points out a
severe failure of the PS approach, but makes it even more difficult for a
high normalization (e.g. based on COBE) CDM spectrum to satisfy constraints
on cluster scales.  On the other hand, the reduced negative evolution in
the number of hot clusters could alleviate the discrepancy between the
expected abundances in critical universes and incoming data, which instead
seem to favour low $\Omega_0$ (Henry 1997).  We are extending the analysis
to other CDM spectra in order to address this point.

\acknowledgements

We thank our collaborators: A. Babul, T. Quinn and J. Stadel, for allowing  
us to show these results prior of publication.  We thank C. Baugh for 
his comments.

\end{document}